\documentclass[12pt]{article}
\usepackage{amssymb,graphicx}
\usepackage{epstopdf}
\usepackage{amsmath,amsfonts}
\usepackage{epsfig}
\usepackage{xcolor}
\usepackage[utf8]{inputenc}
\usepackage{appendix}

\makeatletter
\newcommand{\mathleft}{\@fleqntrue\@mathmargin0pt}
\newcommand{\mathcenter}{\@fleqnfalse}
\makeatother

\begin{document}
\begin{flushright}
INR-TH-2020-038
\end{flushright}
\vspace{10pt}
\begin{center}
  {\LARGE \bf  Horndeski genesis:\\[0.3cm]
    consistency of classical theory}

  \vspace{20pt}
Y. Ageeva$^{a,b,c,}$\footnote[1]{{\bf email:} ageeva@inr.ac.ru}, P. Petrov$^{a,b,}$\footnote[2]{{\bf email:} petrov@ms2.inr.ac.ru, pavelkpetrov@mail.ru} and V. Rubakov$^{a,b,}$\footnote[3]{{\bf email:} rubakov@ms2.inr.ac.ru}\\
\vspace{15pt}
  $^a$\textit{
Department of Particle Physics and Cosmology, Physics Faculty, M.V. Lomonosov Moscow State University, 119991 Moscow, Russia
  }\\
\vspace{5pt}
$^b$\textit{
Institute for Nuclear Research of
         the Russian Academy of Sciences,\\  60th October Anniversary
  Prospect, 7a, 117312 Moscow, Russia}\\
\vspace{5pt}
$^c$\textit{
  Institute for Theoretical and Mathematical Physics,
M.V. Lomonosov Moscow State University, 119991 Moscow, Russia
  }
    \end{center}
    \vspace{5pt}

\begin{abstract}

Genesis within
the Horndeski theory is one of possible scenarios for the start of
the Universe. 
In this model, the absence of instabilities is obtained at the expense of
the property that coefficients, serving as 
effective Planck masses, vanish
in the asymptotics $t\rightarrow -\infty$, which
signalizes the danger of strong coupling and inconsistency of the classical
treatment.
We investigate this problem
in a specific model and extend the analysis of
cubic action for 
perturbations (arXiv:2003.01202)
to arbitrary order.
Our study is based on power counting and
dimensional analysis of the higher order terms.
We derive the latter, find characteristic strong coupling
energy scales and obtain the conditions for the validity
of the classical description. Curiously,  we find
that the strongest condition is the same as that obtained in
already examined cubic case. 


\end{abstract}

\section{Introduction}
\label{intro}
Genesis \cite{Creminelli:2010ba, Creminelli:2012my, Hinterbichler:2012fr, Hinterbichler:2012yn, Nishi:2015pta, Kobayashi:2015gga} 
is an interesting non-singular alternative to, or completion of
inflationary 
cosmology.
In this scenario, the Universe starts its expansion
from static, Minkowski space-time at zero energy density.
At the initial stage,
energy density builds up and the
Hubble parameter grows.
This requires the violation
of the null energy condition (NEC),
see Ref.\cite{Rubakov:2014jja} for a review of models with NEC-violation.
Models with unusual matter
which violates the NEC or null convergence condition~\cite{Tipler:1978zz} 
often suffer from pathological behavior because of
various kinds of instabilities. 
It was noticed, however,
that in Horndeski theory, the NEC can be violated in a stable way. 
Horndeski theory \cite{Horndeski:1974wa, Fairlie:1991qe, Luty:2003vm, Nicolis:2004qq, Nicolis:2008in, Deffayet:2010qz, Kobayashi:2010cm, Padilla:2012dx, Kobayashi:2019hrl} is a scalar-tensor modification of gravity, with the Lagrangian
containing second derivatives of the scalar field and yet with the
second-order equations of motion. Stable NEC-violation is
insufficient for constructing a complete cosmological model, though:
it was shown in
Refs.~\cite{Libanov:2016kfc,Kobayashi:2016xpl}
that the absence of instabilities
imposes strong constraints on Horndeski genesis.
Nevertheless, there is an example of the
Lagrangian \cite{Kobayashi:2016xpl} which yields stable genesis at the level of
classical field theory and linear perturbations. A potential drawback
of the model of Ref.~\cite{Kobayashi:2016xpl} is that 
``effective Planck masses''
vanish in the
asymptotic past, which may lead to the strong coupling problem and
make the classical treatment irrelevant\footnote{
  It has been shown~\cite{Cai:2016thi,Creminelli:2016zwa,Cai:2017dyi,Kolevatov:2017voe,Mironov:2019qjt} that
  another way to get
  around the constraints of Refs.~\cite{Libanov:2016kfc,Kobayashi:2016xpl}
  is to make use of beyond
  Horndeski~\cite{Zumalacarregui:2013pma, Gleyzes:2014dya} or DHOST
  theories~\cite{Langlois:2015cwa,Langlois:2018dxi}.}.

%
In Refs.~\cite{Ageeva:2018rhc,Ageeva:2020gti},
the strong coupling problem in the model of Ref.~\cite{Kobayashi:2016xpl}
has been addressed at the level of cubic action for perturbations.
By making use of the  dimensional analysis, it has been shown that
there exists a region in the parameter space where the classical
field theory treatment is legitimate despite the fact that
``effective Planck masses''
vanish as $t\to -\infty$.
In Ref.~\cite{Petrov:2020vlq} the strong coupling problem was examined in
another model of genesis, involving vector galileon.

The purpose of this paper is to extend the analysis of
Refs.~\cite{Ageeva:2018rhc,Ageeva:2020gti} to all orders of perturbation theory
and figure out whether the same conclusion holds: the classical field
theory is adequate for describing the Horndeski genesis of
ref.~\cite{Kobayashi:2016xpl} in a fairly large region of
the parameter space.

Let us remind how the strong coupling problem arises in the genesis model of 
Ref.~\cite{Kobayashi:2016xpl}. Let
$h_{ij}$ and $\zeta$
denote tensor and scalar metric
perturbations about spatially flat 
Friedmann--Lema\^itre--Robertson--Walker (FLRW) background
solution in the unitary gauge $\delta\phi = 0$,
where $\phi$ is Horndeski scalar field. The unconstrained
quadratic action for these perturbations has the general form
 \begin{eqnarray}
 \label{second_order_unconstr}
        \mathcal{ S}_{(2)}=\int dt d^3x N_0 a^3\left[
        \mathcal{ G}_S
        \frac{\dot\zeta^2}{N_0^2}
        -\frac{\mathcal{ F}_S}{a^2}
        \zeta_{,i}\zeta_{,i} + \mathcal{ G}_T\frac{\dot h_{ij}^2}{8N_0^2}-\frac{\mathcal{ F}_T}{8a^2}
        h_{ij,k}h_{ij,k} 
        \right] \; ,
 \end{eqnarray}
 where
 $\mathcal{ F}_S$, $\mathcal{ G}_S$, $\mathcal{ F}_T$, and $\mathcal{ G}_T$
are functions of cosmic time $t$, $a(t)$ is the scale factor, and $N_0$ is the background lapse function.
To avoid ghost and gradient instabilities
one requires that the coefficients
satisfy
    \begin{align*}
        \mathcal{ F}_S, \mathcal{ G}_S, \mathcal{ F}_T, \mathcal{ G}_T>0 \; .
    \end{align*}
In the case of genesis, the background
is nonsingular: $a(t) \to 1$ as $t\to - \infty$, while $N_0=1$.
Thus, if the functions
$\mathcal{ F}_S$, $\mathcal{ G}_S$, $\mathcal{ F}_T$, and $\mathcal{ G}_T$
are bounded from below by a strictly positive number,
then the integral
    \begin{equation*}
        \int_{-\infty}^{t} a(t) [\mathcal{ F}_T (t)+\mathcal{ F}_S(t)] dt
    \end{equation*}
    is divergent at the lower limit of integration.  The no-go theorem
of Refs.~\cite{Libanov:2016kfc,Kobayashi:2016xpl}
    states that in these
circumstances, there is a gradient or ghost instability at some
stage of the cosmological evolution.

The model of Ref.~\cite{Kobayashi:2016xpl} makes use of the
observation that
this no-go theorem no longer holds if 
$\mathcal{ F}_T \to 0$,
$\mathcal{ F}_S \to 0$,  $\mathcal{ G}_T \to 0$ and
$\mathcal{ G}_S \to 0$ as $t \to -\infty$. Their asymptotics
are~\cite{Kobayashi:2016xpl}
    \begin{equation}
    \label{second_order_coeff_asym}
        \mathcal G_{T}\propto(-t)^{-2\mu},\;\;
        \mathcal F_{T}\propto(-t)^{-2\mu},\;\; 
        \mathcal G_{S}\propto(-t)^{-2\mu+\delta},\;\;
        \mathcal F_{S}\propto(-t)^{-2\mu+\delta},\;\;
    \end{equation}
where $\mu$ and $\delta$ are model parameters with $2\mu > 1 + \delta$ and $\delta > 0$.
At the same time, this behavior
implies that one may
encounter  strong coupling regime in the asymptotic past,
since the 
coefficients
  of quadratic action for metric perturbations, which serve as
 effective Planck masses, 
tend to zero
as $t\to -\infty$.

However, it has been pointed out in Ref.~\cite{Ageeva:2018rhc}
that the fact that $\mathcal{ F}_T$,
$\mathcal{ F}_S$,  $\mathcal{ G}_T$, and
$\mathcal{ G}_S$ tend to zero as $t \to -\infty$ does not necessarily mean
that the classical field theory is not applicable for describing
the evolution of the background in this asymptotics. To see what is going on,
one has to estimate 
%
the actual strong
coupling energy scale $E_{strong}$
by studying cubic and higher-order interactions. 
The classical analysis is legitimate  
for Horndeski genesis if 
the energy scale  $E_{class}$ characteristic of
the classical
background evolution is lower than
$E_{strong}$, 
    \begin{equation*}
        E_{class} \ll  E_{strong}.
    \end{equation*}
    Here the classical energy scale is the inverse time scale
    of the background evolution; for power-law behavior of the
    background one has
     \begin{equation*}
        E_{class} \sim |t|^{-1}\; .
     \end{equation*}

     As we mentioned above, for the model of  Ref.~\cite{Kobayashi:2016xpl}
     this  program has been carried out in
     Refs.~\cite{Ageeva:2018rhc,Ageeva:2020gti} at the level of cubic terms
     in the action for perturbations, and here we consider all orders.
     We
     make use of naive dimensional analysis based on
 power counting and find
     the strongest constraints on the parameters of the
     Lagrangian at each order in nonlinearity.
     When doing so, we disregard at all steps
     any
     possible cancellations 
     and, in particular,
     do not take care of numerical coefficients. The cancellations,
     if any, can only enlarge the region of the parameter space where
     the strong coupling problem does not occur.
%
%
We find that  
the strongest
constraint is the same as that
coming from  the cubic action of
the scalar perturbation   studied in detail 
in Ref. \cite{Ageeva:2020gti}, i.e., higher order nonlinearities
do not add anything new insofar as the strong coupling issue is concerned.

This paper is organized
as follows. In Sec.~\ref{general_frame}  
we describe the model
and 
derive the general form of conditions for the absence of
strong coupling.
In Sec.~\ref{s_c_analys} we introduce
our technique
based on naive power counting and dimensional analysis of 
higher order action. Using this technique
we find  
the strongest  constraint on model parameters
that ensures that the strong coupling energy scale is parametrically
above the classical energy scale in the asymptotics $t \to -\infty$.
We discuss our results in Sec. \ref{concl}. 
In Appendix \ref{app:lagr_all_var} we present the expansion of
the action in all metric perturbations; in Appendix  \ref{app:constr}
we express non-dynamical variables through
$\zeta$ and $h_{ij}$ by solving the constraint equations, still
within the perturbation theory and our power counting technique.
Finally, Appendix \ref{app:unconstr_lagr} is dedicated to the derivation 
of the unconstrained action.

\section{Generalities}
\label{general_frame}

\subsection{The model}
We study the  genesis model of Ref.~\cite{Kobayashi:2016xpl}
which belongs to
a simple subclass of   Horndeski theories.
The covariant form of the action
for this
subclass  is
    \begin{equation}
        \label{action}
        \mathcal{S} = \int d^3x dt  \sqrt{-g}\mathcal{L},
    \end{equation}
where
    \begin{eqnarray}
    \cal L&=&G_2(\phi, X)-G_3(\phi, X)\Box \phi+ G_4(\phi)R,
    \label{Hor_L}\\
        X &=& -\frac{1}{2}g^{\mu\nu}\partial_{\mu}\phi\partial_{\nu}\phi,
    \nonumber
    \end{eqnarray}
    $R$ is the Ricci scalar, $\Box \phi = g^{\mu\nu} \nabla_\mu \nabla_\nu \phi$
    and
$G_{2,3,4}$ are some functions of their variables. 
We use  mostly plus metric signature $(-,+,+,+)$ and work in natural units, 
i.e. $c=\hbar=G=1$. 

Instead of the covariant form, it is 
convenient for our purposes  to use the 
Arnowitt--Deser--Misner (ADM) decomposition\footnote{The
  way to convert one formalism to
  another can be found in
  Refs.~\cite{Gleyzes:2014dya, Gleyzes:2013ooa, Fasiello:2014aqa}}
  of the Lagrangian \eqref{Hor_L}: 
    \begin{align}
        \mathcal{L} =  A_2 (t, N) + A_3 (t, N) K
        +  A_4 (t, N) (K^2 - K^{i}_{j}K^{j}_{i}) + B_4 (t, N)\;^{(3)} R \text{,}
        \label{ADM_Hor_L}
    \end{align}
where $\phi=const$ hypersurfaces are taken to be constant time hypersurfaces.
When
  it comes to perturbations, 
  the latter property means that 
  we choose the
unitary gauge,
\[
\delta \phi = 0.
\]
The general form of metric is
   \begin{equation}
   \label{metric}
     ds^2=-N^2 dt^2 +  
        \gamma_{ij}\left( dx^i+N^i dt\right)\left(dx^j+N^j dt\right) ,
    \end{equation}
   where $\gamma_{ij}$ is the spatial metric.

The extrinsic curvature  and the spatial Ricci 
tensor
are
    \begin{align*}
        K_{ij} &\equiv\frac{1}{2N}\big(\dot\gamma_{ij} -\;^{(3)}\nabla_{i}N_{j}-\;^{(3)}\nabla_{j}N_{i}\big),  \\
        ^{(3)}R_{ij}  &\equiv \partial_{k}\;^{(3)}\Gamma^{k}_{ij}-\partial_{i}\;^{(3)}\Gamma^{k}_{kj}+\;^{(3)}\Gamma^{k}_{lk}\;^{(3)}\Gamma^{l}_{ij}-\;^{(3)}\Gamma^{k}_{li}\;^{(3)}\Gamma^{l}_{jk},
    \end{align*}
while $K= \gamma^{ij}K_{ij}$, $^{(3)} R = \gamma^{ij} \phantom{0}^{(3)} R_{ij}$.
Finally, 
$\sqrt{-g} \equiv N\sqrt{\gamma}$ in action  \eqref{action}, 
where $\gamma \equiv \text{det}(^{(3)}\gamma_{ij})$.

We study concrete  Horndeski theory of
Ref.~\cite{Kobayashi:2016xpl}, in which the
Lagrangian functions are specified as follows:
    \begin{subequations}
    \label{adm_func_lagr}
    \begin{align}
        &A_2 =  f^{-2\mu -2 -\delta} a_2 (N) \text{,} \\ 
        &A_3 =  f^{-2\mu -1 -\delta} a_3 (N) \text{,} \\
        & B_4 = - A_4 = f^{-2\mu} \text{,}
        \label{B_4}
    \end{align}
    \end{subequations}
where $\mu$ and $\delta$ are constant parameters\footnote{Note 
that in Refs.\cite{Kobayashi:2016xpl,Ageeva:2020gti}, $\mu$ parameter 
was denoted by $\alpha$. We save the notation $\alpha$ for a
metric variable.}, the same as in
\eqref{second_order_coeff_asym}, 
and $f(t)$ is some function of time such that
    \begin{equation*}
        f \propto -t,  \hspace{5mm} t \rightarrow - \infty  \text{.}
    \end{equation*}
It was shown in  Ref.~\cite{Kobayashi:2016xpl}
that one gets around the no-go theorem  
and builds
genesis cosmology by choosing
    \begin{equation}
    \label{Kob_conditions}
        2\mu > 1 + \delta \; ,  \hspace{5mm}  \delta > 0 \; .
    \end{equation}
We use this choice in what follows.
    The functions $a_2$ and $a_3$ entering
    \eqref{adm_func_lagr} are given by
    \begin{subequations}
    \begin{align}
      &a_2(N) = -\frac{1}{N^2} + \frac{1}{3N^4} \text{,}
        \label{smol_a2}\\
        &a_3(N) = \frac{1}{4N^3}\text{.}
    \label{smol_a3}
    \end{align}
    \end{subequations} 
 The asymptotics of the
background genesis 
solution \cite{Kobayashi:2016xpl}
is
    \begin{equation*}
        a \propto 1 + \frac{1}{\delta (-t)^\delta}, \quad N_0 \to 1 , \quad \text{as} \quad t\to-\infty,
    \end{equation*}
    where $a(t)$ is the scale factor and
    $N_0$ is the background value of lapse function $N$. 
The Hubble parameter is
$H= \dot{a}/(N_0a)$ and equals
    \begin{equation*}
        H \propto \frac{1}{(-t)^{1+\delta}} \; .
    \end{equation*}
Wherever possible, we use the asymptotic values $a=N_0=1$.


In this paper we concentrate on the analysis of
  perturbations. The ADM decomposition of the 
metric \eqref{metric}, perturbations included,
reads
    \begin{align*}
        N &=N_0(1+\alpha),\nonumber\\
        N_{i} &=\partial_{i}\beta + N^T_i, \; \text{where} \; \partial_i N^{Ti}=0,\nonumber\\
        \gamma_{ij} &=a^{2}\Big(e^{2\zeta}(e^{h})_{ij} + \partial_i\partial_j Y + \partial_i W^T_j + \partial_j W^T_i\Big) \; .
    \end{align*}
    We fix residual gauge freedom by setting $Y = 0$ and $W^T_i = 0$, so the
    spatial part of metric reads
    \begin{align*}
        \gamma_{ij} &=a^{2}e^{2\zeta}(e^{h})_{ij}
    \end{align*}
with
    \begin{align*}
        (e^h)_{ij} &=\delta_{ij}+h_{ij}+\frac{1}{2}h_{ik}h_{kj}+\frac{1}{6}h_{ik}h_{kl}h_{lj}+\cdots, \quad  h_{ii}=0, \; 
        \partial_{i}h_{ij}=0 \; .
    \end{align*}
    Variables
      $\alpha$, $\beta$ and $N^T$ enter the action without temporal
    derivatives;
    the dynamical degrees of freedom are $\zeta$
    and transverse traceless $h_{ij}$, i.e., scalar and tensor perturbations.
    

\subsection{Sketch of the analysis}
\label{sub_sketch}

The purpose  of the further discussion in this Section
is to present the general scheme for deriving 
the strong coupling energy scales 
coming from interaction terms in the action.
We adopt the most straightforward approach and perform our analysis
by making use of  the \textit{unconstrained} action written in terms
of variables $\zeta$ and $h$ (we often omit indices in $h_{ij}$).

For power-counting purposes, we disregard all numerical coefficients,
make use of \eqref{ADM_Hor_L}, \eqref{adm_func_lagr}
and schematically 
write the asymptotic (large $-t$) expression for the
integrand in the action
as follows:
    \begin{align}
        \label{asympt_adm_lagr}
        \sqrt{g}\mathcal{L}
        \propto (-t)^{-2\mu}
        \Big[(-t)^{-2 -\delta}a_2 (N) + (-t)^{-1 -\delta} a_3 (N) K
        +  (K^2 - K^{i}_{j}K^{j}_{i}) + \;^{(3)} R\Big] N \sqrt{\gamma} \text{.}
    \end{align}
    By varying this action  with respect 
    to non-dynamical variables
    $\alpha$, $\beta$ and $N^T_i$ one obtains constraint equations,
    then solves these equations for  $\alpha$, $\beta$ and $N^T_i$,
    plugs them back into the action 
    and obtains the unconstrained action for $\zeta$ and $h$.
    Importantly, the parameter $\mu$
    enters the overall prefactor in \eqref{asympt_adm_lagr}
    only, so
    the constraint equations and hence expressions for
    $\alpha$, $\beta$ and $N^T_i$ are independent of $\mu$, while the
    unconstrained action has the prefactor   $(-t)^{-2\mu}$.

    We then expand the unconstrained action in $\zeta$ and $h_{ij}$.
    Quadratic part is given by \eqref{second_order_unconstr}.
    In accordance to the above discussion,
a higher order term of  $p$-th order in scalar $\zeta$ 
and $q$-th order in tensor $h_{ij}$ in the integrand of the 
unconstrained
action ($p+q \geq 3$) has the following schematic
form:
    \begin{equation}
        \label{term__schemat_lagr}
        (\sqrt{-g}\mathcal{L})_{(pq)} \propto (-t)^{-2\mu}\sum_{l} (-t)^{d_l}\cdot (\partial_t)^{a_l} \cdot
        ( \partial_i)^{b_l} \cdot \zeta^p \cdot h^q
        \; ,
    \end{equation}
    where subscript $(pq)$ refers to orders in $\zeta$ and $h$
    (no multiplication of $p$ and $q$),  $l$ labels different
    types of terms,
    $a_l$ and $b_l$ are the numbers of temporal
    and spatial derivatives (each acting on either $\zeta$
    or $h$), respectively, and
    $d_l$ are linear in $\delta$. In our dimensional analysis we
    discard the numerical coefficients in \eqref{term__schemat_lagr}.
    An example of the term \eqref{term__schemat_lagr} is
    the cubic action in the scalar sector ($p=3$, $q=0$), written in  
    Refs.~\cite{DeFelice:2011zh,Gao:2011qe,Ageeva:2020gti}; in that case,
    the sum in  \eqref{term__schemat_lagr} has 17 terms with different
    numbers of derivatives and/or different time-dependent coefficients. 
    As the dimensionality
    of temporal and spatial derivatives is the same, 
for our dimensional analysis 
we rewrite \eqref{term__schemat_lagr} as
    \begin{equation}
        (\sqrt{-g}\mathcal{L})_{(pq)} \propto (-t)^{-2\mu}\sum_{l} (-t)^{d_l}\cdot (\partial)^{c_l} \cdot  \zeta^p \cdot h^q  
      \;,
      \label{term_schemat_lagr_c_l}
    \end{equation}
where we introduce general derivative operator $\partial$ 
and count the number of these operators with $c_l \equiv a_l + b_l$. Clearly,
the number of terms in the sum in \eqref{term_schemat_lagr_c_l} is smaller than in
 \eqref{term__schemat_lagr}.

The next step is the canonical normalization of 
perturbations $\zeta$ and $h_{ij}$. 
The explicit form of the
unconstrained quadratic action
\eqref{second_order_unconstr} and
asymptotic behavior of coefficients \eqref{second_order_coeff_asym}
immediately give the canonically normalized fields
    \begin{equation*}
        (h_{ij})_{(c)}  \propto \sqrt{\mathcal{G}_T} h_{ij} \propto (-t)^{-\mu} h_{ij}, 
    \end{equation*}
and
    \begin{equation*}
        \zeta_{(c)}\propto \sqrt{\mathcal{G}_S} \zeta \propto(-t)^{-\mu+\frac{\delta}{2}}\zeta \; .
    \end{equation*}
The fact that the coefficients here tend
to zero  as  $t \to -\infty$ (due to the restrictions \eqref{Kob_conditions} 
imposed on the Lagrangian parameters) 
is crucial, as it signalizes
possible 
strong coupling regime at early times. 
In terms of the canonically normalized fields, we have
\begin{align}
        \label{new_form_lagr_term}
        (\sqrt{-g}\mathcal{L})_{(pq)} &\propto(-t)^{-2\mu}\sum_{l} (-t)^{d_l}\cdot (\partial)^{c_l} \cdot \mathcal{G}_S^{-p/2} \cdot \zeta_{(c)}^p \cdot \mathcal{G}_T^{-q/2} \cdot h_{(c)}^q \nonumber \\ 
        &\equiv \sum_{l} \Lambda_l\cdot (\partial)^{c_l} \cdot \zeta_{(c)}^p \cdot  h_{(c)}^q,
\end{align}
where
    \begin{equation*}
        \Lambda_l \equiv (-t)^{-2\mu+d_l}\mathcal{G}_S^{-p/2} \mathcal{G}_T^{-q/2} 
        =  (-t)^{-2\mu+d_l+p(\mu-\delta/2) +q\mu}.
    \end{equation*}

Now, we make use of  dimensional analysis 
and find strong coupling
energy scale $E_l$ associated with each of the terms in the
sum in \eqref{new_form_lagr_term}.
The dimension of canonically
normalized fields is $[\zeta_{(c)}]= [(h_{ij})_{(c)}] = 1$, 
while $[\sqrt{-g}\mathcal{L}] = 4$.
Thus, the dimension of $\Lambda_l$ is
    \begin{equation*}
      [\Lambda_l] = [\mathcal{L}] - [\zeta_{(c)}^p] -  [h_{(c)}^q] - [\partial^{c_l}] 
        = 4-p-q - c_l.
    \end{equation*}
Note that only terms with $4-c_l-p-q<0$ potentially 
lead to strong coupling. The strong coupling energy scale $E_l$
is 
    \begin{equation*}
        E_{l} \propto \Lambda_l^{- \frac{1}{c_l+p+q-4}}
        \propto (-t)^{-\frac{-2\mu+d_l+p(\mu-\delta/2) +q\mu}{c_l+p+q-4}} \; .
    \end{equation*}
    The requirement of the legitimacy of the
    classical treatment  is  
    $ E_{class} \ll  E_l$ for any $l$. The classical energy scale
    is inferred from $\dot{H}/H \propto (-t)^{-1}$ (the scale
$H \propto (-t)^{-1-\delta}$ is lower), so we have $E_{class} \propto (-t)^{-1}$.
    Thus, by requiring $E_{class} \ll E_l$,
    we obtain that a given monomial of $(pq)$-order yields the 
following condition imposed on the Lagrangian parameters $\mu$ and $\delta$
(which must obey  \eqref{Kob_conditions}):
    \begin{equation}
    \label{original_form_SC_cond}
    -2\mu+d_l+p(\mu-\delta/2) +q\mu < c_l+p+q-4 ,
    \end{equation}
with $p+q\geq3$, 
$4-c_l-p-q<0$.
We rewrite \eqref{original_form_SC_cond} as
    \begin{equation}
    \label{No_SC_general}
         \mu < 1 + \frac{p\delta}{2(p+q-2)} - \frac{(d_l - c_l) + 2}{p+q-2},
    \end{equation}
and see 
that the most dangerous of  $(pq)$-terms
are those with the largest difference $(d_l - c_l)$.
So, at given $(pq)$-order, we have to find the term in
$(\sqrt{-g}\mathcal{L})_{(pq)}$ with the largest
$d_l-c_l$, and then obtain the smallest right hand side
of \eqref{No_SC_general} among all $p$ and $q$ (with $p+q \geq 3$).
This will give the smallest region of healthy Lagrangian parameters.

We implement this procedure in Sec.~\ref{s_c_analys};
details of calculations are given in Appendices A, B and C.

\section{Implementation of the procedure}
\label{s_c_analys}

\subsection{Simplifications}
\label{sub_simplifications}
Explicitly
evaluating the perturbative expansion of the original action,
solving constraints and obtaining the unconstrained action to
arbitrary order appears notoriously difficult. However, we make
a number of simplifications. We will find in the end that the strongest
constraint on the model parameters
 is the same as that
coming from the cubic terms
which have been already
analyzed  in detail~\cite{Ageeva:2020gti}.
Therefore, all these simplifications do not modify our final result.
 Our simplifications are as follows.

1) As mentioned above, we discard all numerical factors thus
neglecting any possible cancellations.
 Also, 
we do not keep track of the tensor structure of various terms in the
cubic and higher order action.

2) In accordance with the above discussion, 
for given $p$ and $q$ we keep only those monomials entering
\eqref{term_schemat_lagr_c_l} in
the unconstrained action,  which
have the largest value of $(d_l - c_l)$. We write the value of
$(d_l - c_l)$ as a superscript in front of the expression involving the
fields, i.e., we employ the notation
    \begin{equation}
        (-t)^d (\partial)^c \zeta^p h^q = ^{(d-c)}\zeta^p h^q
        \label{new_index}
    \end{equation}
and do not distinguish monomials with different $d$ and $c$ but
the same $(d-c)$.

3) We do the same in the cubic and higher order terms in the
original action involving all variables
$\alpha$, $\beta$, $N^T$, $\zeta$ and $h$.
To see why this is correct, we
write the constraint equations:
\begin{subequations}
  \label{general_var_set}
\begin{align}
  \frac{\delta(\sqrt{-g}\mathcal{L})^{(2)}}{\delta \alpha}+\frac{\delta(\sqrt{-g}\mathcal{L})^{H.O.}}{\delta \alpha}=0,
\label{general_var_alpha}
  \\
  \frac{\delta(\sqrt{-g}\mathcal{L})^{(2)}}{\delta \beta}+\frac{\delta(\sqrt{-g}\mathcal{L})^{H.O.}}{\delta \beta}=0,
\label{general_var_beta}
  \\
        \frac{\delta(\sqrt{-g}\mathcal{L})^{(2)}}{\delta N^T}
        +\frac{\delta(\sqrt{-g}\mathcal{L})^{H.O.}}{\delta N^T}=0,
\label{general_var_N_T}
\end{align}
\end{subequations}
 where $(\sqrt{-g}\mathcal{L})^{(2)}$ is the quadratic part
   of the original integrand in action, which is
   known explicitly (and whose structure will be given below),
   and $(\sqrt{-g}\mathcal{L})^{H.O.}$ contains cubic and higher order terms.
   A general term in the latter is (to
     simplify formulas here,
   we do not write the overall factor
   $t^{-2\mu}$ in  $(\sqrt{-g}\mathcal{L})$, see \eqref{asympt_adm_lagr},
   \eqref{term_schemat_lagr_c_l})
   \begin{equation}
        (-t)^D (\partial)^C \alpha^{m_\alpha} \beta^{m_\beta} (N^{T})^{m_{N^T}}
        \zeta^{m_\zeta} h^{m_h} 
        \label{general_term_all_pert}
   \end{equation}
   with positive integer   $m_\alpha, \dots, m_h$.
   Let us compare effects of terms with the same set of parameters
   $(m_\alpha, \dots, m_h)$ and different $C$ and $D$.
   Upon solving the constraint equations, one finds
   $\alpha (\zeta, h)$,   $\beta (\zeta, h)$,   $N^T (\zeta, h)$, again as
   series in $\zeta$, $h$ and their derivatives, with coefficients
   depending on $t$. By substituting them back into \eqref{general_term_all_pert} one finds
   the term
  \begin{equation*}
    (-t)^D (\partial)^C \alpha^{m_\alpha} (\zeta, h)
    \beta^{m_\beta} (\zeta, h)  (N^{T}(\zeta, h))^{m_{N^T}}
   \zeta^{m_\zeta} h^{m_h} \; .
  \end{equation*}
  which is a linear combination of expressions \eqref{new_index}.
  The largest value of $(d-c)$ for given $(p,q)$ is obtained
  for the largest value of $(D-C)$.
  Also, order by order in perturbation theory,
    the largest contributions
  to $\alpha$, $\beta$ and $N^T$ (in the sense of the largest
  $(d-c)$ for given $(p,q)$ in the unconstrained action)
  come from the terms  with the largest $(D-C)$.
  So, for  given $(m_\alpha, \dots, m_h)$
  we keep only the terms in  the original
  $(\sqrt{-g}\mathcal{L})^{H.O.}$
  with the largest $(D-C)$ and, in analogy with
  \eqref{new_index}, use the notation
  \begin{equation}
   (-t)^D (\partial)^C \alpha^{m_\alpha} \beta^{m_\beta}  (N^{T})^{m_{N^T}}
  \zeta^{m_\zeta} h^{m_h} =  ^{(D-C)}\alpha^{m_\alpha} \beta^{m_\beta} (N^{T})^{m_{N^T}}
  \zeta^{m_\zeta} h^{m_h} \; 
  \label{general_term_all_var_new_index}
  \end{equation}
  and do not distinguish terms with the same $(D-C)$ but different
  $D$ and $C$.

  4) Yet another
  simplification is that we replace some terms in the original
  $(\sqrt{-g}\mathcal{L})^{H.O.}$, which have the form
\eqref{general_term_all_pert},
  with new ones with larger
  $(D-C)$. This can only strengthen the constraint on the model
  parameters (but in fact it does not).
  Concretely, we make the following replacements
in
  $(\sqrt{-g}\mathcal{L})^{H.O.}$ (we use the notation \eqref{general_term_all_var_new_index}):
\begin{subequations}
    \label{simplific_1}
\begin{align}
        &^{-2-\delta}(1+\alpha+\ldots) \to ^{-2}(1+\alpha+\ldots), \\ 
    &^{-3-\delta}\beta(1+\alpha+\ldots) \to ^{-3}\beta(1+\alpha+\ldots),
    \\
    &^{-2-\delta}(1+\alpha+\ldots)\zeta N^T
    \to ^{-2}(1+\alpha+\ldots)\zeta  N^T \; .
      \end{align}
\end{subequations}
We note in passing that
       similar replacements in quadratic part  $(\sqrt{-g}\mathcal{L})^{(2)}$
       (more precisely, in the part bilinear in non-dynamical variables)
     would be impossible, since they would have an effect of
     erroneously weakening the constraint on model parameters.
      Finally, we can add arbitrary extra terms
      to   $(\sqrt{-g}\mathcal{L})^{H.O.}$; again,
      this can only strengthen the constraint on the model
  parameters (but in fact it does not). We use this observation to replace
     \begin{align}
    \label{simplific_2}
        (1+\alpha) \to (1+\alpha+\alpha^2+\ldots)
    \end{align}
     when expanding the term
     $\sqrt{-g}B_4 ^{(3)} R$ in \eqref{ADM_Hor_L} (and only at that point).
     These replacements
     simplify the calculations considerably.

\subsection{Dominant terms in the action}
\label{sub_lagr_all_var}

With the above notations and
simplifications, the dominant terms in the original
action have fairly transparent form.
 We recall that there is
  the overall factor
   $t^{-2\mu}$ in  $(\sqrt{-g}\mathcal{L})$, see \eqref{asympt_adm_lagr},
  \eqref{term_schemat_lagr_c_l}, and that it does not contribute to superscripts
  $(D-C)$ and $(d-c)$, which are independent of $\mu$.
  Therefore, it is convenient to write the expressions
  for
  $t^{2\mu}\sqrt{-g}\mathcal{L}$ instead of $\sqrt{-g}\mathcal{L}$.
   The calculation is described
in  Appendix \ref{app:lagr_all_var} and gives
    \begin{align}
    \label{simple_gL}
        &t^{2\mu}\sqrt{-g}\mathcal{L} \supset t^{2\mu}(\sqrt{-g}\mathcal{L})^{(2)}+ \Big\{(1+\alpha+\alpha^2+\ldots)\Big[\;^{-2}(1+\zeta+\zeta^2+\ldots) 
        \nonumber \\ 
        &+ ^{-2}(1+\zeta+\zeta^2+\ldots)(h^2+h^3+\ldots) \nonumber \\ 
        &+ ^{-3}\beta(1+\zeta+\zeta^2+\ldots)(1+h+h^2+\ldots) 
        \nonumber \\ 
        &+ ^{-2}N^T(\zeta+\zeta^2+\ldots) +^{-2}N^T(1+\zeta+\zeta^2+\ldots)(h+h^2+\ldots) \nonumber \\
        &+ ^{-4}\beta^2 (1+\zeta+\zeta^2+\ldots)(1+h+h^2+\ldots) + ^{-2}(N^T)^2(1+\zeta+\zeta^2+\ldots)(1+h+h^2+\ldots) \nonumber \\ 
        &+^{-3}\beta N^T (1+\zeta+\zeta^2+\ldots)(1+h+h^2+\ldots)\Big]\Big\}\;^{H.O.} \; ,
    \end{align}
    where notation
    $\{\ldots\}^{H.O.}$ means that linear and quadratic parts are omitted.
We write  $\supset$ sign instead of equality or proportionality signs, 
since we make replacements \eqref{simplific_1} and \eqref{simplific_2},
proceed with the naive analysis and do not care about 
numerical coefficients and possible cancellations. 

The exact second order integrand  $(\sqrt{-g}\mathcal{L})^{(2)}$
was evaluated in  Ref.~\cite{Kobayashi:2011nu}, except for the
term involving $N^T$. The latter term is straightforwardly
calculated and has the form
 $(\partial_iN^T_j)^2$ 
due to transversality of $N_i^T$. We again omit numerical
coefficients and write
    \begin{align*}
       t^{2\mu}(\sqrt{-g}\mathcal{ L})^{(2)}&\propto 
        \dot\zeta^2+ \zeta_{,i}\zeta_{,i}
        +(-t)^{-\delta-2} \alpha^2 \nonumber \\
        &+(-t)^{-\delta-1}\alpha\beta_{,ii}
        + \dot \zeta\beta_{,ii}
        +(-t)^{-\delta-1} \alpha\dot\zeta
        +\alpha\zeta_{,ii} \nonumber  \\ 
        &+ \dot h_{ij}^2+ h_{ij,k}h_{ij,k} \nonumber \\
        &+ (\partial_i N_j^T)^2.
    \end{align*}
    Let us compare  terms $(-t)^{-\delta-1} \alpha\dot\zeta$ and
    $\alpha\zeta_{,ii}$. In our notations, these are written
    as  $^{- \delta -2}\alpha \zeta$ and
    $^{-2}\alpha \zeta$, respectively. Since $\delta > 0$,
    the latter term dominates over the former one both in the constraint
    equation \eqref{general_var_alpha}
    and in the unconstrained action. So, we neglect the former term.
    After that, the second order part of the integrand of the action
    is written as follows:
    \begin{align}
        \label{second_ord_new_form}
        t^{2\mu} (\sqrt{-g}\mathcal{L})^{(2)}\supset \;^{-2}\zeta^2+\;^{-2-\delta}\alpha^{2}
        +\;^{-3-\delta}\alpha\beta+\;^{-3}\zeta\beta+\;^{-2}\zeta\alpha + \;^{-2}h^2 + \;^{-2}(N^{T})^{2} \; .
    \end{align}
    Note that superscripts  of bilinears of non-dynamical variables here
    (terms with $\alpha^2$, $\alpha \beta$ and $(N^T)^2$) have two-fold role.
    On the one hand, they determine the structure of the linear terms
    in the constraint equations \eqref{general_var_set} and hence enter the
    perturbative solution to these equations with flipped signs. As an example,
    to the linear order,
    the constraint equation \eqref{general_var_beta} has the form
    $^{-3-\delta}\alpha+\;^{-3}\zeta =0$ and gives
    $\alpha \propto \; ^{\delta}\zeta$. On the other hand, these superscripts
    appear with their original signs in the expressions for the parts of
    the unconstrained
    action obtained by plugging the solutions to constraints
    $\alpha (\zeta, h)$,   $\beta (\zeta, h)$ and   $N^T (\zeta, h)$
    back into \eqref{second_ord_new_form}.

\subsection{Solutions for $\alpha$, $\beta$ and $N^T_i$} 
\label{sub_constr}
We now solve the constraint equations \eqref{general_var_set} and find
$\alpha$, 
$\beta$ and $N^T_i$, still within our power-counting approach.
To the linear order, we use the action
\eqref{second_ord_new_form}, and find that 
the solutions are (see Appendix \ref{app:constr} for details)
\begin{subequations}
  \label{section_3_linear_constr}
    \begin{align}
        \alpha_{(pq)} &= \;^{\delta}\zeta, \label{alp_1} \\
        \beta_{(pq)} &= \;^{\delta+1}\zeta, \label{beta_1} \\ 
        N^T_{(pq)} &= 0 \label{N_1}, \;\;\;\;\;\; p+q=1 \;.
    \end{align}
    \end{subequations}
Note that linear order solutions involve $\zeta$ only, 
which should be the case
since $h_{ij}$ is traceless and transverse tensor and
thus there is no way to construct a linear scalar structure out of it.
Also, the solution for $N^T_i$
does not have a linear term.

The solutions to quadratic order are obtained by plugging
\eqref{section_3_linear_constr} in the  cubic
part of the action
\eqref{simple_gL} and using the result in \eqref{general_var_set}.
We do this calculation in  Appendix \ref{app:constr}.
Keeping the dominant terms only (with the largest $(d-c)$
in each term) we obtain
\begin{subequations}
  \label{section_3_quadr_constr}
    \begin{align}
        \alpha_{(pq)} &= ^{\delta}h^2 + ^{3\delta}\zeta^2 
        + ^{2\delta}\zeta h, \label{alp_2} \\
        \beta_{(pq)} &= \;^{\delta+1}h^2 + ^{3\delta+1}\zeta^2 + ^{2\delta +1}\zeta h, \label{beta_2}\\ 
        N^T_{(pq)} &= h^2 + ^{2\delta}\zeta^2 + ^{\delta }\zeta h \label{N_2},
        \;\;\;\;\;\; \;\;\;\;\;\; p+q=2 \;.
    \end{align}
    \end{subequations}
    We obtain higher order terms in Appendix \ref{app:constr} by
    induction and get
    \begin{subequations}
      \label{section_3_n-1_constr_set}
    \begin{align}
    \label{alpha_constr}
        \alpha_{(pq)} &= ^{(2p+q-1)\delta}\zeta^p h^q  ,
    \\
    \label{beta_constr}
        \beta_{(pq)} &= ^{(2p+q-1)\delta+1}\zeta^p h^q ,
\\
    \label{N_T_constr}
    (N^T)_{(pq)} &= ^{(2p+q-2)\delta}\zeta^p h^q  \; ,
    \;\;\;\;\;\;\;\;\;\;\;\;\; p+q \geq 2 \; .
    \end{align}
    \end{subequations}
    The complete solutions are linear combinations of \eqref{section_3_linear_constr}
    and \eqref{section_3_n-1_constr_set} with all $p$ and $q$ such that $p+q \geq 2$.

\subsection{Unconstrained action} 
\label{sub_unconstr_lagr}

Now, we substitute the solutions \eqref{section_3_n-1_constr_set}
  into the second and 
  higher order parts of the action integrand
  \eqref{simple_gL}.
  We keep terms with maximum value of $(d-c)$ 
  for each combination with fixed $\zeta^p h^q$ in the
  unconstrained action of cubic and higher order.
  The details of the
  calculation are given in Appendix~\ref{app:unconstr_lagr},
  and here we quote the results.
  The contribution coming from the second, explicit higher order
  term in   \eqref{simple_gL} is
    \begin{align}
        \label{final_unconstr_act_m3}
        t^{2\mu} (\sqrt{-g}\mathcal{L})^{H.O.}
        \supset \sum_{p+q \geq 3}\;^{(2p+q-3)\delta -2}\zeta^{p}h^{q} \; .
    \end{align}
    The quadratic action
    \eqref{second_ord_new_form} gives the following contribution:
    \begin{align}
    \label{final_unconstr_act_m2}
      t^{2\mu}(\sqrt{-g}\mathcal{L})^{(2)}\supset
    \sum_{2p+q \geq 4 \,, \; p+q\geq 3}\;^{(2p+q-3)\delta -2}
        \zeta^{p}h^{q} \; ,
    \end{align}
    where the condition $p+q \geq 3$ reflects the fact that we
    are interested in cubic and higher order terms.
    We see that the structure of the leading terms in the unconstrained
    action is not particularly complicated.

    \subsection{Constraint on $\mu$ and $\delta$}

    Now, we recall that a term $^{(d-c)} \zeta^p h^q$ ($p+q \geq 3$)
      yields a constraint  \eqref{No_SC_general} on the model parameters,
      which we reproduce here:
    \begin{equation*}
      \mu < 1 + \frac{p\delta}{2(p+q-2)} -
      \frac{(d - c) + 2}{p+q-2}, \quad p+q\geq3.
    \end{equation*}
Each term in the expression
    \eqref{final_unconstr_act_m3} 
     has $d-c = (2p+q-3)\delta -2$ and therefore gives the constraint
     	\begin{equation}
	  \mu + \delta + \frac{\delta(p-2)}{2(p+q-2)} <1 \; ,
          \;\;\;\;\; p+q \geq 3 \; .
\label{final_condition_H_O}
        \end{equation}
        It is straightforward to see that the strongest of these constraints
        comes from the terms with $q=0$, $p \geq 3$ (which includes
        cubic order in the
        scalar sector), and reads
    \begin{equation}
        \mu + \frac{3}{2}\delta<1 \; .
\label{final_strongest_condition}
    \end{equation}
    This constraint coincides with the result of Ref.~\cite{Ageeva:2020gti}
    obtained at cubic order.

The cubic and higher order terms in
    $(\sqrt{-g}\mathcal{L})^{(2)}$, given by \eqref{final_unconstr_act_m2},
also have $(d-c) = (2p+q-3)\delta -2$. The constraints on parameters
have the same form as \eqref{final_condition_H_O},
 but now the range of $p$ and $q$ is
 $2p+q \geq 4$, $p+q \geq 3$.
  The strongest of these constraints again comes from terms with
 $q=0$, $p \geq 3$ and has the form \eqref{final_strongest_condition}.
  Thus, the model with parameters obeying \eqref{final_strongest_condition}
  (together with $2\mu > 1+\delta$ and $\delta >0$, see
  \eqref{Kob_conditions})
 is free of strong coupling problem as long as the validity of
 classical description of genesis is concerned.

\section{Conclusion}
\label{concl}
To summarize, the model we studied in this paper
admits a consistent classical field theory description of
the early genesis stage, provided its parameters are chosen in the
range
    \begin{align*}
    \label{healthy_region}
        &2\mu > 1 + \delta>1, \nonumber\\
        &\mu + \frac{3}{2}\delta<1.
    \end{align*}
    This genesis epoch is peculiar, as it begins, as $t\to - \infty$,
    at zero ``effective Planck masses'', which appears
    necessary in Horndeski theories (unlike in
    their generalizations)
    for avoiding instabilities during the entire
    evolution. Yet the quantum strong coupling
    energy scale $E_{strong}$
    stays well above the energy scale of classical evolution
    $E_{class} \sim t^{-1}$, to the extent that
    \[
    \frac{E_{strong} (t)}{E_{class} (t)} \to \infty \;\;\;\;
    \mbox{as}\;\; t \to - \infty \; .
    \]
    This is because the interaction terms in the action for perturbations
    vanish rapidly enough in early-time asymptotics.

    Clearly, the model studied in this paper is just an example
    of a consistent theory of so peculiar beginning of the Universe.
    Its advantage is that it is simple enough to allow for a
    reasonably straightforward analysis of the strong coupling issue, as
    we demonstrated in this paper. The power counting techniques we
    introduced may possibly
    be extended to more complicated models.
    
    It would be interesting to examine various ways to incorporate
    such a genesis model 
    into a full cosmological scenario, i.e. invent and
    study a healthy transition
to the next stage like inflation  or straight to the
conventional hot epoch. First steps in this direction
have been made already~\cite{Nishi:2016wty}.

\section*{Acknowledgments} 
We are grateful to D. Ageev for comments 
on the manuscript and to O. Evseev for 
helpful discussions.  This work has been
supported by Russian Science Foundation Grant No. 19-12-00393.

\appendix

\section{Expansion of $\sqrt{-g}\mathcal{L}$ 
  in $\alpha$, $\beta$, $\zeta$, $N^T_i$ and $h_{ij}$}
\label{app:lagr_all_var}
\numberwithin{equation}{section}

\setcounter{equation}{0}

In this Appendix we expand  $\sqrt{-g}\mathcal{L}$
in metric perturbations. We discard all numerical factors and keep
only the dominant terms, as described in Secs.~\ref{sub_sketch} and
\ref{sub_simplifications}. For this reason we use the symbol
$\supset$ instead of equality sign.

The quadratic action is known explicitly, so we concentrate on
cubic and higher order terms.
We begin with the expression for the three-dimensional Christoffel symbol
       $ ^{(3)}\Gamma_{ij}^{k} = \frac{1}{2}\gamma^{ka}(\gamma_{ai,j} + \gamma_{aj,i} - \gamma_{ij,a})$.
We substitute $\gamma_{ij} =a^{2}e^{2\zeta}(e^{h})_{ij}$ here
and evaluate the the derivative of tensor exponent:
    \begin{equation*}
      (e^h)_{ij,\; l} = h_{ij,\; l}+\frac{1}{2}h_{ik, \;l}\;h_{kj}
      + \frac{1}{2}h_{ik}\;h_{kj,\; l} +\cdots \; .
    \end{equation*}
We obtain
    \begin{align*}
        \;^{(3)}\Gamma_{ij}^{k}  &\supset (e^{-h})^{ka}e^{-2\zeta}[\big((e^{h})_{ai}e^{2\zeta}\partial_{j}\zeta+(e^{h})_{aj}e^{2\zeta}\partial_{i}\zeta - (e^{h})_{ij}e^{2\zeta}\partial_{a}\zeta\big)\\
        +&\big(e^{2\zeta}\partial_{j}h_{ai}+e^{2\zeta}\partial_{i}h_{aj} - e^{2\zeta}\partial_{a}h_{ij}\big)
        \\ 
        +&\big(e^{2\zeta}h_{ab}\partial_{j}h_{bi}+e^{2\zeta}h_{bi}\partial_{j}h_{ab} +  e^{2\zeta}h_{ab}\partial_{i}h_{bj}+e^{2\zeta}h_{bj}\partial_{i}h_{ab} - e^{2\zeta}h_{ib}\partial_{a}h_{bj}-e^{2\zeta}h_{bj}\partial_{a}h_{ib}\big)
        \\  
        +&\ldots ],
    \end{align*}
    where dots stand for higher order terms in $h_{ij}$. Since
    $(e^h)_{ik}(e^{-h})^{kj} = \delta^j_i$,
we write Christoffel symbols schematically as
    \begin{align*}
        \;^{(3)}\Gamma_{ij}^{k} &\supset \big(\delta_{ki}\partial_{j}\zeta+\delta_{kj}\partial_{i}\zeta
        -\delta_{ij}\partial_{k}\zeta\big) \nonumber\\
        &+\big(\partial_{i}h_{kj}+\partial_{j}h_{ik}-
        \partial_{k}h_{ij}\big)+\big(\partial h^{2}+\partial h^{3}+\ldots\big)_{kij} \; .
    \end{align*}
where
$\big(\partial h^{2}+\partial h^{3}+...\big)_{kij}$ 
includes terms  
$h_{ab}\partial_{j}h_{bi}$, $h_{bi}\partial_{j}h_{ab}$, etc.
We keep the tensor structure of the linear terms here
  and in appropriate places below, since
we will encounter cancellations associated with it.
Hereafter
all spatial indices in final expressions
are lower ones, so that
there are no hidden metric factors like $e^{2\zeta}$ or $(e^{h})_{ij}$.

The Lagrangian \eqref{ADM_Hor_L} involves 
extrinsic curvature which we write as
\[
K_{ij} = \frac{E_{ij}}{N} \; ,
\]
where
    \begin{align*}
        E_{ij} &=\frac{1}{2}\Big(\dot\gamma_{ij} -\;^{(3)}\nabla_{i}N_{j}-\;^{(3)}\nabla_{j}N_{i}\Big),
    \end{align*}
with $^{(3)}\nabla_i N_j \equiv \partial_i N_{j} - ^{(3)}\Gamma_{ij}^{k}N_{k}$, 
and $N_i = \partial_i \beta + N^T_i$.
The term $\dot{\gamma}_{ij}$ reads
    \begin{align*}
        \dot{\gamma}_{ij} = \frac{\partial}{\partial t} \Big(a^2 e^{2\zeta} (e^{h})_{ij}\Big) \supset H e^{2\zeta} (e^{h})_{ij} + \dot{\zeta} e^{2\zeta} (e^{h})_{ij} + e^{2\zeta} (\dot{h}_{ij} + h_{ik}\dot{h}_{kj} + h_{kj}\dot{h}_{ik} + \ldots),
    \end{align*}
    where dots again denote
higher order terms in $h_{ij}$.
We do not expand $e^{2\zeta}$ in all terms  here 
and $(e^{h})_{ij}$ in the first two terms, 
since the next step is the contraction $E^{i}_{j} = \gamma^{ik}E_{kj}$ 
where some of $e^{2\zeta}$ and $(e^{h})_{ij}$  cancel out.
In notations of Sec.~\ref{sub_simplifications} we have
    \begin{align*}
        \dot{\gamma}_{ij} \supset (H  + ^{-1}{\zeta}) (e^{h})_{ij} e^{2\zeta} + ^{-1}(h + h^2 + h^3 + \ldots)_{ij} e^{2\zeta}.
    \end{align*}
    We make similar steps for the terms
    $^{(3)}\nabla_iN_j + ^{(3)}\nabla_jN_i$ and obtain
    \begin{align*}
        E_{j}^{i} &\supset (H+\;^{-1}\zeta)\delta^{i}_{j}+
        \;^{-1}(h+h^{2}+\ldots)_{ij}  \nonumber \\ 
        &+(1+\zeta+\zeta^2+\ldots)\Big((1+h+h^2+\ldots)_{ik}\;\partial_{k}\partial_{j}\beta + (1+h+h^2 + \ldots)_{lj}\;^{(3)}\Gamma_{il}^{k}\;\partial_{k}\beta \nonumber \\ 
        &+(1+h+h^2+\ldots)_{ik}\;(\partial_{k}N^{T}_{j}+
        \partial_{j}N^{T}_{k})
        +(1+h+h^2+\ldots)_{lj}\;^{(3)}\Gamma_{il}^{k}\;N^{T}_{k}\Big) \; .
    \end{align*}
    When
      evaluating the trace $E\equiv E^i_i$ and contraction
$E_i^j E^j_i$,
    we will 
encounter cancellations due to the properties
$h_{ii} = \partial_i h_{ij} = 0$ 
    and $\partial_i N^{Ti} =0$.

    We are ready to expand various terms in  $(-t)^{2\mu}\sqrt{-g}\mathcal{L}$
    in metric perturbations (the
    reason for including the factor   $(-t)^{2\mu}$
    is explained in Sec.~\ref{sub_lagr_all_var}).
    The factor $Ne^{3\zeta}$ in the left hand sides below
    comes from $\sqrt{-g}$. We obtain the following.
\begin{itemize}
\item $\sqrt{-g} A_2$. This term is straightforwardly calculated by
   expanding in $\alpha$ the function $a_2(N)$ given by
\eqref{smol_a2}:
    \begin{equation*}
      (-t)^{2\mu}N e^{3\zeta} A_2
      \supset ^{-2-\delta}(1+\alpha+\alpha^2+\ldots)(1+\zeta+\zeta^2 +\ldots) \; .
    \end{equation*}
  \item $\sqrt{-g} A_3 K$.  Making
    use of expansion of function $a_3(N)$ given by \eqref{smol_a3}, we write 
    \begin{align*}
        E \equiv K \cdot N &\supset H + ^{-1}\zeta + ^{-1}(h^2+\ldots) \\
        &+ ^{-2}\beta(1+\zeta+\ldots)(1+h+\ldots) \\ 
        &+^{-1}N^T(1+\zeta+\ldots)(h+\ldots)
        + ^{-1}N^T(\zeta+\ldots), 
    \end{align*}
and find
    \begin{align*}
      &  (-t)^{2\mu}N e^{3\zeta} A_3 K
      \supset ^{-1-\delta}(1+\alpha+\ldots)\Big[\;^{-1-\delta} 1 + ^{-1}(\zeta+\ldots) + ^{-1}(1+\zeta+\ldots)(h^2+\ldots) \\ 
        &+ ^{-2}\beta(1+\zeta+\ldots)(1+h+\ldots) +^{-1}N^T(1+\zeta+\ldots)(h+\ldots)
        + ^{-1}N^T(\zeta+\ldots)\Big],
    \end{align*}
where the term $^{-1-\delta}1$ comes from the Hubble parameter 
$H\propto (-t)^{-1-\delta}$. 
\item $\sqrt{-g} A_4 (K^2 - K^{i}_{j}K^{j}_{i})$. A straightforward
  calculation gives
    \begin{align*}
        E^2 &\supset ^{-2-2\delta}1 + ^{-2}\zeta^2 + ^{-2-\delta}(h^2+h^3)+ ^{-2}(h^4+\ldots) + ^{-2-\delta}\zeta + ^{-2}\zeta(h^2+\ldots) \\ \nonumber
        &+ ^{-4}\beta^2 (1+\zeta+\ldots)(1+h+\ldots) \\ \nonumber
        &+ ^{-2}(N^T)^2(h^2+\ldots)+^{-2}(N^T)^2(\zeta+\ldots)(h+\ldots)+^{-2}(N^T)^2(\zeta^2+\ldots) \\ \nonumber
        &+^{-3}\beta(\zeta+\ldots)(1+h+\ldots) + ^{-3-\delta}\beta + ^{-3}\beta(1+\zeta+\ldots)(h^2+\ldots) \\ \nonumber
        &+ ^{-2-\delta}N^T \zeta + ^{-2}N^T(\zeta^2+\ldots)
        \\ \nonumber
        &+ ^{-2-\delta}N^{T}(h+h^{2}) +^{-2}N^T(h^3+\ldots) +^{-2}N^T(\zeta+\ldots)(h+\ldots) \\ \nonumber
        &+ ^{-3}\beta N^T(h+\ldots) + ^{-3}\beta N^T (\zeta+\ldots) + ^{-3}\beta N^T (\zeta+\ldots)(h+\ldots),
    \end{align*}
    \begin{align*}
        E^{i}_{j}E^{j}_{i} &\supset ^{-2-2\delta}1 + ^{-2}\zeta^2 +  ^{-2}(h^2+\ldots) + ^{-2-\delta}\zeta + ^{-2}\zeta(h^2+\ldots) \\ \nonumber
        &+ ^{-4}\beta^2 (1+\zeta+\ldots)(1+h+\ldots) + ^{-2}(N^T)^2(1+\zeta+\ldots)(1+h+\ldots) \\ \nonumber
        &+^{-3}\beta(1+\zeta+\ldots)(h+\ldots) + ^{-3-\delta}\beta + ^{-3}\beta(\zeta+\ldots) \\ \nonumber
        & +^{-2-\delta}N^T \zeta + ^{-2}N^T(\zeta^2+\ldots) +^{-2}N^T(1+\zeta+\ldots)(h+\ldots) \\ \nonumber
        &+ ^{-3}\beta N^T (1+\zeta+\ldots)(1+h+\ldots) \; .
    \end{align*}
    There is some difference between $E^2$ and $E^{i}_{j}E^{j}_{i}$. 
    In particular, $E^2$ contains
        $^{-2-\delta}(h^2+h^3)+ ^{-2}(h^4+\ldots)$, while
$E^{i}_{j}E^{j}_{i}$ includes another structure
        $^{-2}(h^2+\ldots)$.
    This happens due to the fact that $h_{ij}$ is
    traceless and 
transverse. Together, the two expressions read
    \begin{align*}
        E^2 - E^{i}_{j}E^{j}_{i} &\supset\;^{-2-2\delta}1 + \;^{-2}\zeta^2 + \; ^{-2}(h^2+\ldots) +\; ^{-2-\delta}\zeta +\; ^{-2}\zeta(h^2+\ldots) \\ \nonumber
        &+\; ^{-4}\beta^2 (1+\zeta+\ldots)(1+h+\ldots) +\; ^{-2}(N^T)^2(1+\zeta+\ldots)(1+h+\ldots) \\ \nonumber
        &+\;^{-3}\beta(1+\zeta+\ldots)(h+\ldots) +\; ^{-3-\delta}\beta +\; ^{-3}\beta(\zeta+\ldots) \\ \nonumber
        & +\;^{-2-\delta}N^T \zeta +\; ^{-2}N^T(\zeta^2+\ldots) +\;^{-2}N^T(1+\zeta+\ldots)(h+\ldots) \\ \nonumber
        &+\; ^{-3}\beta N^T (1+\zeta+\ldots)(1+h+\ldots) \; ,
    \end{align*}
and we obtain 
    \begin{align*}
        (-t)^{2\mu}N e^{3\zeta} A_4 (K^2 &- K^{i}_{j}K^{j}_{i}) \supset (1+\alpha+\ldots)\Big[\;^{-2-2\delta}1 \\ \nonumber
        &+ \;^{-2-\delta}\zeta + ^{-2}(\zeta^2+\ldots) +  ^{-2}(1+\zeta+\ldots)(h^2+\ldots)  \\ \nonumber
        &+ \;^{-4}\beta^2 (1+\zeta+\ldots)(1+h+\ldots)  \\ \nonumber
        &+ ^{-2}(N^T)^2(1+\zeta+\ldots)(1+h+\ldots) \\ \nonumber
        &+\;^{-3}\beta(1+\zeta+\ldots)(h+\ldots) + ^{-3-\delta}\beta + ^{-3}\beta(\zeta+\ldots) \\ \nonumber
        &+ \; ^{-2-\delta}N^T \zeta + ^{-2}N^T(\zeta^2+\ldots) +^{-2}N^T(1+\zeta+\ldots)(h+\ldots) \\ \nonumber
        &+ ^{-3}\beta N^T (1+\zeta+\ldots)(1+h+\ldots)\Big]\; .
    \end{align*}
  \item $\sqrt{-g} B_4 \; ^{(3)}R$. We again make use of
    the fact that $h_{ij}$ is
    traceless and 
transverse and find
    \begin{align*}
          (-t)^{2\mu}N e^{3\zeta} B_4 \; ^{(3)}R \supset (1+\alpha)\Big[\;^{-2}(1+\zeta+\ldots)(h^2+\ldots) + ^{-2}(\zeta+\ldots)\Big] \; .
    \end{align*}
    Note that unlike other terms,
    this term contains
the factor $(1+\alpha)$ instead of the
full series $(1+\alpha+\alpha^2+\ldots)$. The reason is
that both  $B_4 = B_4(t)$ and  $^{(3)}R$ are independent of $N$.
\end{itemize}
Collecting all terms together, we find
    \begin{align*}
    &(-t)^{2\mu} (\sqrt{-g}\mathcal{L})^{H.O.} \supset
    \Big[\;^{-2}(1+\alpha)(\zeta+\ldots) + ^{-2-\delta}(\alpha^2+\ldots)\zeta \nonumber \\ 
        &+ (1+\alpha+\ldots)\big(\;^{-2-\delta}1 + ^{-3-\delta}\beta + ^{-2-\delta}\zeta N^T + ^{-2}(\zeta^2+\ldots) + ^{-2}(1+\zeta+\ldots)(h^2+\ldots) \nonumber \\ 
        &+^{-3}\beta(1+\zeta+\ldots)(h+\ldots) + ^{-3}\beta(\zeta+\ldots) + ^{-2}N^T(\zeta^2+\ldots) +^{-2}N^T(1+\zeta+\ldots)(h+\ldots) \nonumber \\ 
        &+ ^{-4}\beta^2 (1+\zeta+\ldots)(1+h+\ldots) + ^{-2}(N^T)^2(1+\zeta+\ldots)(1+h+\ldots) \nonumber \\ 
    &+^{-3}\beta N^T (1+\zeta+\ldots)(1+h+\ldots)\big)\Big]^{H.O.} \; .
    \end{align*}
    This expression is simplified by making use of \eqref{simplific_1},
    i.e., removing $\delta$ from all superscripts, and using
    \eqref{simplific_2} in the first term in square brackets.
    The result is given by \eqref{simple_gL}.
    
\section{Solution to constraint equations}
\label{app:constr}
\numberwithin{equation}{section}

\setcounter{equation}{0}

In this Appendix we solve the constraint equations \eqref{general_var_set}
and find non-dynamical variables  $\alpha$, $\beta$ and $N^T_i$
in the form of series in  $\zeta$ and $h_{ij}$.

We begin with Eq.~\eqref{general_var_beta}. The calculation of the
variations of \eqref{simple_gL} and \eqref{second_ord_new_form}
with respect to $\beta$ is straightforward and gives
(hereafter we do not write the overall factor $(-t)^{2\mu}$) 
    \begin{align}
        \frac{\delta (\sqrt{-g}\mathcal{L})^{H.O.}}{\delta \beta} &= \;^{-3}(\alpha^2+\ldots)+^{-3}(h^2+\ldots) + ^{-3}(\zeta^2+\ldots) \nonumber \\ 
        &+ ^{-3}(\alpha+\ldots)(h+\ldots) + ^{-3}(\alpha+\ldots)(\zeta+\ldots) + ^{-3}(\zeta+\ldots)(h+\ldots) \nonumber \\ 
        &+ ^{-3}(\alpha+\ldots)(h+\ldots)(\zeta+\ldots) \nonumber \\ 
        &+ (^{-3}N^T + ^{-4}\beta)\big[(\alpha+\ldots) + (\zeta+\ldots) + (h+\ldots) \nonumber \\ 
        &+ (h+\ldots)(\alpha+\ldots) + (h+\ldots)(\zeta+\ldots) + (\alpha+\ldots)(\zeta+\ldots) \nonumber \\ 
        &+ (\alpha+\ldots)(\zeta+\ldots)(h+\ldots) \big], \label{alpha_constr_m3}
         \\ 
        \frac{\delta (\sqrt{-g}\mathcal{L})^{(2)}}{\delta \beta} &= \;^{-3-\delta}\alpha + ^{-3}\zeta \label{alpha_constr_m2},
    \end{align}
    To the linear order, the relevant equation is \eqref{alpha_constr_m2},
    and we immediately get
   \begin{equation}
\alpha_{(pq)}=\;^{\delta}\zeta \;,\;\;\; p+q=1.
\label{app:alpha_linear_constr}
   \end{equation} 
Next, we  vary the action with respect to $\alpha$ 
and write
    \begin{align}
        \frac{\delta (\sqrt{-g}\mathcal{L})^{H.O.}}{\delta \alpha} &= \;^{-2}(\alpha^2+\ldots) + ^{-2}(\zeta^2+\ldots) + ^{-2}(\alpha+\ldots)(\zeta+\ldots) \nonumber \\ 
        &+ ^{-2}(1+\alpha+\ldots)(1+\zeta+\ldots)(h^2+\ldots) \nonumber \\ 
        &+^{-3}\beta(h+\ldots)+ ^{-3}\beta(\zeta+\ldots) +^{-3}\beta(\alpha+\ldots) \nonumber \\
        &+^{-3}\beta(\alpha+\ldots)(\zeta+\ldots) + ^{-3}\beta(h+\ldots)(\zeta+\ldots) +^{-3}\beta(\alpha+\ldots)(h+\ldots)\nonumber \\
        &+^{-3}\beta(\alpha+\ldots)(\zeta+\ldots)(h+\ldots) \nonumber \\ 
        &+^{-2}N^T(1+\alpha+\ldots)(\zeta+\ldots) +^{-2}N^T(1+\alpha+\ldots)(1+\zeta+\ldots)(h+\ldots) \nonumber \\ 
        &+^{-4}\beta^2(1+\alpha+\ldots)(1+\zeta+\ldots)(1+h+\ldots) \nonumber \\ 
        &+^{-2}(N^{T})^{2}(1+\alpha+\ldots)(1+\zeta+\ldots)(1+h+\ldots) \nonumber \\  &+ ^{-3}\beta N^T(1+\alpha+\ldots)(1+\zeta+\ldots)(1+h+\ldots), \label{beta_constr_m3} \\ 
        \frac{\delta (\sqrt{-g}\mathcal{L})^{(2)}}{\delta \alpha} &= \;^{-2-\delta}\alpha+
        \;^{-3-\delta}\beta + ^{-2}\zeta. \label{beta_constr_m2}
    \end{align}
    Again considering linear order, and using \eqref{beta_constr_m2}
    and \eqref{app:alpha_linear_constr}, 
we find 
        \begin{equation}
        \beta_{(pq)} = \;^{\delta+1}\zeta\; ,\;\;\; p+q=1.
\label{app:beta_linear_constr}
        \end{equation}   

Finally, we turn to the variation of the action with respect to $N^T$:
    \begin{align}
        \frac{\delta (\sqrt{-g}\mathcal{L})^{H.O.}}{\delta N^T} &= \;^{-2}(\zeta^2+\ldots)+^{-2}(\alpha+\ldots)(\zeta + \ldots) \nonumber \\ &+(^{-2}h + ^{-3}\beta + ^{-2}N^T)\big[(h+\ldots) + (\zeta+\ldots) + (\alpha+\ldots) \nonumber \\ 
        &+ (\alpha+\ldots)(h+\ldots) + (\zeta+\ldots)(h+\ldots) + (\alpha+\ldots)(\zeta+\ldots)\nonumber \\ 
        &+(\alpha+\ldots)(\zeta+\ldots)(h+\ldots)\big], \label{N_T_constr_m3}
        \\ 
        \frac{\delta (\sqrt{-g}\mathcal{L})^{(2)}}{\delta N^T} &= \;^{-2}N^T, \label{N_T_constr_m2}
    \end{align}
    so that to the linear order we have
    \begin{equation}
        N^T_{(pq)} = 0, \;\;p+q=1.
        \label{app:N_T_linear_constr}
    \end{equation}
The linear order solution is summarized in \eqref{section_3_linear_constr}.

Obtaining perturbative solution is in principle straightforward:
to find the solution to order $p+q=n$, one writes the unknown
$n$-th order  $\alpha$, $\beta$ and $N^T$ in the  linear parts of the
constraint equations \eqref{alpha_constr_m2}, \eqref{beta_constr_m2},
\eqref{N_T_constr_m2}, 
plugs the known lower order
expressions
for $\alpha$, $\beta$ and $N^T$ in non-linear parts
\eqref{alpha_constr_m3}, \eqref{beta_constr_m3}, \eqref{N_T_constr_m3},
evaluates these parts to $n$-th order and solves the resulting
equations
for $n$-th order variables.
To quadratic order, we use \eqref{app:alpha_linear_constr},
\eqref{app:beta_linear_constr},  \eqref{app:N_T_linear_constr} in quadratic parts
of the constraint equations. As an example,  Eq.~\eqref{general_var_beta}
reads at quadratic order
    \begin{align*}
        &\frac{\delta (\sqrt{-g}\mathcal{L})^{(2)}}{\delta \beta}\Big|_{(pq)}+\frac{\delta (\sqrt{-g}\mathcal{L})^{H.O.}}{\delta \beta}\Big|_{(pq)} = 
        \\
        =&^{-3-\delta}\alpha_{(pq)}
        +  ^{-3+2\delta}\zeta^2 + ^{-3}h^2 + ^{-3}\zeta^2 + ^{-3+\delta}\zeta h + ^{-3+\delta}\zeta^2 + ^{-3}\zeta h = 0,\;\;
        p+q=2 \; .
    \end{align*}
    We keep the dominant terms (with the largest values of superscripts)
    and obtain the second order result
      \begin{equation*}
        \alpha_{(pq)} = \;^{3\delta}\zeta^2 + ^{\delta}h^2 
        + ^{2\delta}\zeta h,\;\;p+q=2.
    \end{equation*}
      Similar procedure is used to find, with known second-order
      $\alpha$, 
            the expression for the second-order $\beta$ from
  Eq.~\eqref{general_var_alpha},     
    \begin{equation*}
      \beta_{(pq)} = \; ^{3\delta+1}\zeta^2
      +^{\delta+1}h^2 + ^{2\delta +1}\zeta h,\;\;p+q=2,
    \end{equation*}
    and, finally, second-order $N^T$ from Eq.~\eqref{general_var_N_T},
    \begin{equation*}
        (N^T)_{(pq)} = \; ^{2\delta}\zeta^2 +h^2 + ^{\delta }\zeta h.
    \end{equation*}
Due to algebraic cancellations 
and the transversality and tracelessness of 
$N^T_{i}$ and $h_{ij}$, some  terms in the expressions
above may possibly vanish.
Nevertheless, we keep all terms,  
since we proceed with
naive analysis and do not take into account these subtleties.
The second order solution is summarized in \eqref{section_3_quadr_constr}.
    
Let us show by induction that the $k$-th order terms in
the solutions to constraint equations are
 \begin{equation}
        \alpha_{(pq)} = \; ^{(2p +q-1)\delta}\zeta^p h^q, \;\;\;\;\;\;p+q =k,
\label{app:alpha_k_order_constr}
 \end{equation}    
    \begin{equation}
    \label{app:beta_to_alpha}
        \beta_{(pq)} = \; ^{1}\alpha_{(pq)},  \;\;\;\;\;\;p+q = k,
    \end{equation}
    \begin{equation}
    \label{app:N_to_beta_and_alpha}
      N^T_{(pq)} = \; ^{-\delta}\alpha_{(pq)} =
      ^{-\delta-1}\beta_{(pq)}, \; \;\;\;\;\;  p+q =k \; ,\;\;\;\;\;\; k>1.
    \end{equation}
    This is the case for $k=2$. Let us assume that this is the case
    for $k \leq n-1$ and show that the same formulas hold for $k=n$.

    The general idea is that the $n$-th order of the non-linear parts of
    the constraint equations involves only   $\alpha_{(pq)}$,
 $\beta_{(pq)}$ and    $N^T_{(pq)}$ at
   orders
    $p+q \leq n-1$, which are known by assumption of induction.
    Thus, the proof reduces to the evaluation of this non-linear parts
    \eqref{alpha_constr_m3}, \eqref{beta_constr_m3},
\eqref{N_T_constr_m3}.
    
One formula we  use in what follows is 
	\begin{align}
	\label{alpha_to_m}
	    &(\alpha^m)_{(pq)} =
	    \Big[\Big(\;^{\delta}\zeta+\big(\;^{3\delta}\zeta^{2}+
	    \;^{2\delta}\zeta h+\;^{\delta}h^{2}\big)+\ldots\nonumber\\
	    &+\big(\;^{(2l-1)\delta}\zeta^{l}+\;^{(2(l-1)+1-1)\delta}\zeta^{l-1}h
	    +\ldots+\;^{(l-1)\delta}h^{l}\big)+\ldots\Big)^{m}\Big]_{(pq)}\nonumber\\
	    &=\;^{(2p+q-m)\delta}\zeta^{p}h^{q},\;\;m>1,
	\end{align}
        which is valid for  $p$ and $q$ obeying $2p+q\geq2m$
(otherwise the left hand side vanishes)
        and,
by assumption of
induction, $p+q\leq n$.
We also  derive another useful formula, where $m>r+s$:
    \begin{align}
    \label{al_z_h_to_m}
        (\alpha^{m-r-s}\zeta^r h^s)_{(pq)} &=  \sum_{p_{1},\;q_{1}}(\alpha^{m-r-s})_{(p_{1}q_{1})}(\zeta^r h^s)_{(p-p_{1} \;q-q_{1})}  \nonumber \\
        &=\sum_{p_{1},\;q_{1}}\big(\delta_{p-p_{1}}^{r}\delta_{q-q_{1}}^{s}\zeta^{p-p_{1}} h^{q-q_{1}}\big)\big(\;^{(2p_{1}+q_{1}-m+r+s)\delta}\zeta^{p_{1}}h^{q_{1}}\big)
        \Big|_{2(m-k-r)\leq 2p_{1}+q_{1};r\leq p;s\leq
         q} \nonumber \\
         &=\;^{2(p-r)+(q-s)-m+s+r}\zeta^{p}h^{q}\Big|_{2(m-s-r)\leq2p-2r+q-s;\;r\leq p;\;s\leq
         q} \nonumber \\
        &=\;^{(2p+q-m-r)\delta}\zeta^p h^q\Big|_{2p+q\geq2m-s;\;\;r\leq p;\;\;s\leq
         q},
    \end{align}
    Here $r \leq p$, $s \leq q$, and $2p+q\geq2m-s$,
otherwise the left hand side vanishes. We also have
$p+q\leq n$ by assumption of induction. 

Using formulas \eqref{alpha_to_m} and \eqref{al_z_h_to_m}, 
we can compare various terms with one and the same structure
$\zeta^ph^q$ in 
constraints
\eqref{alpha_constr_m3}, \eqref{beta_constr_m3},
\eqref{N_T_constr_m3} and keep only ones with the largest value of $(d-c)$.
To this end, we examine each term in \eqref{alpha_constr_m3} one by one.
\begin{itemize}
    \item We begin with the first term in  Eq.~\eqref{alpha_constr_m3}:
    \begin{align}
        ^{-3}(\alpha^2+\alpha^3+\ldots)_{(pq)} =  \underbrace{^{-3 + (2p+q-2)\delta }\zeta^{p}h^{q}}_{\text{from}\; (\alpha^2)_{(pq)}} + \underbrace{^{-3+(2p+q-3)\delta }\zeta^{p}h^{q}}_{\text{from} \; (\alpha^3)_{(pq)}} + \ldots.
\label{app:reduce_series_by_alpha}
    \end{align}
    From \eqref{alpha_to_m} we observe
    that terms with minimum  power
    of $\alpha$ lead to the contributions with the largest value of
    $(d-c)$ for every $(p,q)$. So, the dominant
    term is $(\alpha^2)_{(pq)}$, and we write 
        \begin{align*}
        \;^{-3}(\alpha^2+\alpha^3+\ldots)_{(pq)} = \;^{-3}(\alpha^2)_{(pq)}, \;p+q\leq n ,
        \end{align*}
        where
         \begin{align*}
           (\alpha^2)_{(pq)} =\; ^{(2p+q-2)\delta}\zeta^p h^q, \; 2p+q\geq4 \; .
           \end{align*}
         Note that the term \eqref{app:reduce_series_by_alpha} does not have contributions
         of order $h^2$ and
         $h^3$, because $\alpha$ is at least quadratic in $h$.
        
       \item The next
         two terms $^{-3}(h^2 + \ldots)$
         and $^{-3}(\zeta^2 + \ldots)$ in \eqref{alpha_constr_m3}
         give contributions with
         smaller value of $(d-c)$ as compared to  \eqref{app:reduce_series_by_alpha},
         except for the cases $q=2,\;p=0$ and $q=3,\;p=0$.
           We are not interested in quadratic terms, since we have already
         studied the quadratic order. So,  
from
these two terms, we (temporarily) keep only $^{-3}h^3$.
    
\item The next
  term is $^{-3}(\alpha+\ldots)(h+\ldots)$. Using the
  formula \eqref{al_z_h_to_m}, we observe
  that the term with the minimum $m$ provides
  the contribution with the largest
  $(d-c)$ for every $(p,q)$. Therefore,
%
    \begin{equation*}
        ^{-3}\big((\alpha+\ldots)(h+\ldots)\big)_{(pq)}=
      ^{-3}\Big(\sum_{m=2}\sum^{s=m-1}_{s=1}\alpha^{m-s}h^{s}\Big)_{(pq)}
      =^{-3}(\alpha h)_{(pq)}\;, \;\;\;\;\;\; q\geq 1\; .
    \end{equation*}
   In accordance with
      \eqref{app:alpha_k_order_constr}, we have
    \[
    (\alpha h)_{(pq)} = \; ^{(2p+q-2)\delta}\zeta^p h^q, \; 2p+q\geq3, \; q\geq1,
    \]
    so that for general $(p,q)$  this term is contained in $\alpha^2$,
    coming from Eq.~\eqref{app:reduce_series_by_alpha}, except
    that there are also the
    terms of order $\zeta h$ and $h^3$. 
    The latter is actually
    the dominant cubic term of order $^{-3+\delta}h^3$.

  \item By the same logic as above, we write for the next term
    $\;^{-3}\big((\alpha+\ldots)(\zeta+\ldots)\big)_{(pq)}
    = \;^{-3}(\alpha\zeta)_{(pq)}$.
    However, using  \eqref{al_z_h_to_m} we find that
    \begin{equation*}
        ^{-3}(\alpha\zeta)_{(pq)}=^{-3}(\alpha^{2-1}\zeta^{1})_{(pq)}=
        \;^{-3+(2p+q-3)\delta}\zeta^{p}h^{q},
    \end{equation*}
 and hence this term
gives smaller contribution than \eqref{app:reduce_series_by_alpha}.

\item The next term $\;^{-3}(\zeta+\ldots)(h+\ldots)$
  is obviously subdominant as compared to  \eqref{app:reduce_series_by_alpha}.

\item Again applying the same logic, we write
%
$\;^{-3}(\alpha+\ldots)(\zeta+\ldots)(h+\ldots)=
  \;^{-3}\alpha\zeta h$.
  The contributions due to this term are again subdominant.
\item Finally, there is the set of terms in
  \eqref{alpha_constr_m3} which has the form
  $(\;^{-3}N^T + \;^{-4}\beta)[...]$, where $[\ldots]$ denotes
  $\big[(\alpha+\ldots) + (\zeta+\ldots) + (h+\ldots) + (h+\ldots)(\alpha+\ldots) \\
+ (h+\ldots)(\zeta+\ldots) + (\alpha+\ldots)(\zeta+\ldots) 
 + (\alpha+\ldots)(\zeta+\ldots)(h+\ldots)]$.
  We use the assumption of induction
  \eqref{app:beta_to_alpha} and \eqref{app:N_to_beta_and_alpha}
to write
    \begin{align*}
        \Big((\;^{-3}N^T + \;^{-4}\beta)[...]\Big)_{(pq)}=
        \Big((\;^{-3-\delta}\alpha + \;^{-4+1}\alpha)[...]\Big)_{(pq)}=
        \Big((\;^{-3}\alpha)[...]\Big)_{(pq)}, \;\;p+q\leq n\;.
    \end{align*}
    Then this set of terms
    involves  precisely the same structures as 
    some of the
    terms studied above, so this contribution gives nothing new.
%
 %
\end{itemize}
To summarize, the non-linear term of the constraint equation
\eqref{alpha_constr_m3} has the $(p,q)$ part dominated by
\[
\frac{\delta (\sqrt{-g}\mathcal{L})^{H.O.}}{\delta \beta}\Big|_{(pq)}
\supset
\; ^{-3 + (2p+q-2)\delta}\zeta^p h^q \; .
\]
We recall the form of the linear part, Eq.~\eqref{alpha_constr_m2}, and
write the equation for $\alpha_{(pq)}$ with $p+q=n$,
\[
^{-3-\delta}\alpha_{(pq)} =  ^{-3 + (2p+q-2)\delta}\zeta^p h^q \; .
\]
This gives
\[
\alpha_{(pq)} = ^{(2p+q-1)\delta}\zeta^p h^q \; , \;\;\;\; p+q=n \geq 3 \; ,
\]
as promised.

The analysis of non-linear parts of other constraint equations,
Eqs.~\eqref{beta_constr_m3} and \eqref{N_T_constr_m3}, is essentially
the same as above. We obtain
   \begin{align*}
     \frac{\delta (\sqrt{-g}\mathcal{L})^{H.O.}}{\delta \alpha}\Big|_{(pq)}
        &\supset   ^{-2 + (2p+q-2)\delta}\zeta^p h^q\;, \;\;\;\;p+q\leq n
       \\
       \frac{\delta (\sqrt{-g}\mathcal{L})^{H.O.}}{\delta N^T}\Big|_{(pq)}
       &\supset   ^{-2 + (2p+q-2)\delta}\zeta^p h^q\;, \;\;\;\;p+q\leq n \; .
   \end{align*}
   With linear terms in the constraint equations given by
   \eqref{beta_constr_m2} and \eqref{N_T_constr_m2}, this yields
   \eqref{app:beta_to_alpha} and \eqref{app:N_to_beta_and_alpha}.
   This completes the proof.

\section{Unconstrained action}
\label{app:unconstr_lagr}
\numberwithin{equation}{section}

\setcounter{equation}{0}

Thus, the linear parts of the non-dynamical variables
$\alpha$, $\beta$, $N^T$ are given by \eqref{app:alpha_linear_constr},
\eqref{app:beta_linear_constr},  \eqref{app:N_T_linear_constr}, while higher
order parts are written in \eqref{app:alpha_k_order_constr},
\eqref{app:beta_to_alpha} and \eqref{app:N_to_beta_and_alpha}.
We plug these expressions in the terms \eqref{simple_gL}
and \eqref{second_ord_new_form} of the original action,
obtain the unconstrained action in this way, and extract the
dominant terms. We do this for the higher order action
\eqref{simple_gL} explicitly, while the procedure for the quadratic
action is similar (and simpler). Recall that we are interested
in cubic and higher order terms there.

We firstly express $\beta$ and $N^T_i$ in terms of $\alpha$
using \eqref{app:beta_to_alpha}
and \eqref{app:N_to_beta_and_alpha} and write:
    \begin{align*}
      (-t)^{2\mu} (\sqrt{-g}\mathcal{L})^{H.O.}_{(pq)} &\supset \Big\{(1+\alpha+\ldots)\Big[\;\underbrace{^{-2}(1+\zeta+\ldots)}_{I.} +
        \underbrace{^{-2}(1+\zeta+\ldots)(h^2+\ldots)}_{II.} \\ \nonumber
        &+ \underbrace{^{-2}\alpha(1+\zeta+\ldots)(1+h+\ldots)}_{III.} +
        \underbrace{^{-2}\alpha^2(1+\zeta+\ldots)(1+h+\ldots)}_{IV.}
        \Big]\Big\}^{H.O.}_{(pq)} \;,
    \end{align*}
    where superscript
    $H.O.$ still means that we keep only cubic and higher order
    terms in original variables $\alpha$,  $\beta$, $N^T$, $\zeta$
    and $h$. As an example, there are no terms of order $\alpha$
    and $\alpha^2$.
    
Let us consider each term separately, 
using \eqref{alpha_to_m} and \eqref{al_z_h_to_m} to
extract the dominant contributions. Simple power counting
similar to that employed in Appendix~\ref{app:constr} gives
    \mathleft
    \begin{align*}
        I. \;\; &\{(1+\alpha+\ldots)(1+\zeta+\ldots)\}^{H.O.}_{(pq)} \\ \nonumber
        &\supset (\zeta^3+\ldots)_{(pq)}
        + (\alpha^3+\ldots)_{(pq)} + \big[(\alpha^2 + \ldots)(\zeta+\ldots)\big]_{(pq)} + \big[(\alpha + \ldots)(\zeta^2+\ldots)\big]_{(pq)}\\ &\supset (\alpha^3)_{(pq)},
    \end{align*}
    \mathcenter
Similarly, we find for other three terms:
    \mathleft
    \begin{align*}
        II.\;  \; \{(1+\alpha+\ldots)(1+\zeta+\ldots)(h^2+\ldots)\}^{H.O.}_{(pq)} \supset (\alpha h^2)_{(pq)} + (h^3)_{(pq)}.
    \end{align*}
    \mathcenter
    \mathleft
    \begin{align*}
        III. \; \; \{(\alpha+\ldots)(1+\zeta+\ldots)(1+h+\ldots)\}^{H.O.}_{(pq)}
        \supset (\alpha^3)_{(pq)} + (\alpha h^2)_{(pq)} + (\alpha^2 h)_{(pq)},
    \end{align*}
    \mathcenter
    \mathleft
    \begin{align*}
        IV. \; \; \{(\alpha^2+\ldots)(1+\zeta+\ldots)(1+h+\ldots)\}^{H.O.}_{(pq)}
        \supset (\alpha^3)_{(pq)} + (\alpha^2 h)_{(pq)}.
    \end{align*}
    \mathcenter
By combining these, we  obtain
    \begin{align}
      (-t)^{2\mu} \big(\sqrt{-g}\mathcal{L}\big)^{H.O.}_{(pq)} \supset
      \big(\;^{-2}\alpha^3 + ^{-2}\alpha h^2 + ^{-2}\alpha^2 h +
      ^{-2}h^3\big)_{(pq)},\;\;p+q\geq3.
\label{app:unconstr_action_H_O}
    \end{align}
    Each term in this expression is non-zero in a certain range of
    $p$ and $q$, see Eqs.~\eqref{alpha_to_m} 
    and \eqref{al_z_h_to_m} and remarks below those formulas.
    Namely, we have ($p+q \geq 3$ everywhere)
        \begin{align*}
          (\alpha^{3})_{(pq)}&=
          \;^{(2p+q-3)\delta}\zeta^{p}h^{q}, \;\; \text{with} \; 2p+q\geq6,
          \\
	  (\alpha h^2)_{(pq)}& =\;^{(2p+q-3)\delta}\zeta^{p}h^{q}, \;
          \text{with}\;\; 2p+q\geq4, \;\; q\geq2, 
\\
(\alpha^2 h)_{(pq)} &=  \;^{(2p+q-3)\delta}\zeta^{p}h^{q} \;,\; \;
        \text{with}\;\; 2p+q\geq 5, \;\; q\geq 1\; .
    \end{align*} 
        Still, the linear combination of these terms together
        with explicit $h^3$
        in \eqref{app:unconstr_action_H_O} exhausts all possibilities with $p+q \geq 3$,
        and we obtain finally
    \begin{align*}
        (-t)^{2\mu}(\sqrt{-g}\mathcal{L})_{(pq)}^{H.O.} \supset \;^{-2+(2p+q-3)\delta}\zeta^{p}h^{q}\Big|_{p+q\geq3} \; ,
    \end{align*}
which is our
formula \eqref{final_unconstr_act_m3}. Similar analysis of quadratic
part of the action gives \eqref{final_unconstr_act_m2}.

\end{document}